\documentclass[12pt]{article}
\usepackage{amsfonts,amssymb,epsfig,amsmath}
\usepackage{bbold}
\usepackage{color}
\usepackage{hyperref}
\renewcommand{\text}[1]{#1}

\newcommand{\be}{\begin{equation}}
\newcommand{\ee}{\end{equation}}
\newcommand{\ben}{\begin{displaymath}}
\newcommand{\een}{\end{displaymath}}
\newcommand{\bea}{\begin{eqnarray}}
\newcommand{\eea}{\end{eqnarray}}
\newcommand{\bean}{\begin{eqnarray*}}
\newcommand{\eean}{\end{eqnarray*}}
\newcommand{\nn}{\nonumber \\}
\newcommand{\ba}{\begin{array}}
\newcommand{\ea}{\end{array}}
\newcommand{\bi}{\begin{itemize}}
\newcommand{\ei}{\end{itemize}}

\renewcommand{\theequation}{\arabic{section}.\arabic{equation}}
\def\theequation{\thesection.\arabic{equation}}

\def\l{\lambda}
\def\a{\alpha}

\def\b{\beta}
\def\g{\gamma}
\def\G{\Gamma}
\def\d{\delta}
\def\G{\Gamma}
\def\g{\gamma}
\def\e{\epsilon}
\def\s{\sigma}
\def\e{\epsilon}

\def\m{\mu}
\def\n{\nu}
\def\S{\Sigma}

\newcommand{\cale}{\mbox{${\cal E}$}}

\begin{document}

\makeatletter
\renewcommand{\theequation}{\thesection.\arabic{equation}}
\@addtoreset{equation}{section}
\makeatother

\begin{titlepage}

\vfill
\begin{flushright}
KIAS-P11046\\
QMUL-PH-11-13
\end{flushright}

\vfill

\begin{center}
   \baselineskip=16pt
   {\Large \bf Fermionic T-duality in the pp-wave limit}
   \vskip 2cm
     Ilya Bakhmatov$^\dagger$, Eoin \'O Colg\'ain$^\star$, and Hossein Yavartanoo$^\ddagger$
       \vskip .6cm
             \begin{small}
             {\it $^\dagger$Queen Mary University of London,}\\
			 {\it Centre for Research in String Theory,}\\
			 {\it School of Physics,}\\
			 {\it Mile End Road, London, E1 4NS, England} \\ \vspace{2mm}
      		 \textit{$^\star$Korea Institute for Advanced Study, \\
        	 Seoul 130-722, Korea} \\ \vspace{2mm} 
        	 \textit{$^\ddagger$ Department of Physics, Kyung Hee University, \\ Seoul 130-701, Korea}
             \end{small}\\*[.6cm]
\end{center}

\vfill \begin{center} \textbf{Abstract}\end{center} \begin{quote}
$AdS_5 \times S^5$ and its pp-wave limit are self-dual under transformations involving eight fermionic T-dualities, a property which accounts for symmetries seen in scattering amplitudes in ${\cal N}=4$ super-Yang-Mills. Despite strong evidence for similar symmetries in the amplitudes of three-dimensional ${\cal N} =6$ ABJM theory, a corresponding self-duality in the dual geometry $AdS_4 \times \mathbb{C} \textrm{P}^3$ currently eludes us. Here, working with the type IIA pp-wave limit of $AdS_4 \times  \mathbb{C} \textrm{P}^3$ preserving twenty four supercharges, we show that the pp-wave is self-dual with respect to eight commuting fermionic T-dualities and not the six expected. In addition, we show the same symmetry can be found in a superposition pp-wave and a generic pp-wave with twenty and sixteen unbroken supersymmetries respectively, strongly suggesting that self-duality under fermionic T-duality may be a symmetry of all pp-waves. 

\end{quote} \vfill

\end{titlepage}

\section{Introduction}
Recently,  a better understanding of the amplitude/Wilson loop correspondence \cite{Alday} and the dual superconformal symmetry \cite{Drummond:2007aua,Brandhuber:2007yx} in the context of ${\cal N}=4$ super-Yang-Mills materialised through the realisation \cite{fermTdual, Beisert} that the dual geometry $AdS_5 \times S^5$ could be mapped back to itself under a combination of ordinary bosonic T-dualities \cite{Buscher} and newer fermionic transformations, dubbed fermionic T-dualities. An important prerequisite for performing a fermionic T-duality is the existence of Killing spinors (supersymmetry), and a simple redefinition of sigma-model couplings, results in a dual geometry where the metric is unchanged, though the dilaton and the RR fluxes get modified.

Despite rosy appearances, it is not fair to say that fermionic T-duality is a simple generalisation of bosonic T-duality, as it possesses some unique quirks. Firstly, a requirement on commuting supersymmetries necessitates the need for \textit{complexified} Killing spinors, often leading to backgrounds which are solutions of complexified supergravity. Secondly, unlike Abelian T-duality, it is not a full symmetry of string theory. In this way, the current status of fermionic T-duality is akin to non-Abelian T-duality \cite{de la Ossa:1992vc}, though recent work \cite{Sfetsos:2010uq,Lozano:2011kb} has shown how to generate supergravity solutions there too.
To date, there is a relatively small body of work exploring this fascinating area of fermionic T-duality, some of which may be found here \cite{Adam:2009kt,Fre:2009ki,Hao:2009hw,Dekel:2011qw,ChangYoung:2011rs,Sfetsos:2010xa,Nikolic:2011ps,Grassi:2011zf}. 
 
Another recent interesting development that stokes interest in this area and provides impetus for our study is the steady stream of papers confirming strong evidence for Yangian invariance \cite{Bargheer:2010hn, Lee:2010du}, dual superconformal symmetry \cite{Huang:2010qy,Gang:2010gy}\footnote{A recursion relation for tree-level scattering amplitudes in three-dimensional Chern-Simons-matter theories generalising BCFW \cite{BCFW} was also noted in \cite{Gang:2010gy}.} and amplitude/Wilson loop duality \cite{Bianchi:2011rn,Bianchi:2011dg,Chen:2011vv} in the scattering amplitudes of ABJM \cite{ABJM}. Attempts to find such symmetries in the gravity dual $AdS_4 \times \mathbb{C} \textrm{P}^3$ have so far come up short \cite{Grassi:2009yj,Adam:2010hh,Bakhmatov:2010fp}, leaving quite an intriguing hole in the literature. Given the difficulties experienced there, a goal of this paper is to explore fermionic T-duality, and in particular self-duality under this symmetry, in the simpler setting of the pp-wave limit of ABJM.

Building on the observation of \cite{Bakhmatov:2009be}, that $AdS_5 \times S^5$ remains self-dual under eight fermionic T-dualities in the pp-wave limit (incidentally the same number as were identified in \cite{fermTdual}), we identify commuting fermionic directions allowing a self-dual description of the pp-wave limit \cite{ABJMpp} of $AdS_4 \times \mathbb{C} \textrm{P}^3$. In the pp-wave limit, the geometry considerably simplifies to the extent that one no longer requires bosonic T-dualities for self-duality. This allows a good opportunity to study just the fermionic T-dualities and examine the proposal of \cite{Bargheer:2010hn} from studying the superconformal algebra that six fermionic T-dualities are required. Though the superalgebra will change, it is certainly valid to ask if  a signature of these may be found after one takes the pp-wave limit. 

In contrast to recent null results \cite{Grassi:2009yj,Adam:2010hh,Bakhmatov:2010fp}, in this paper we find something that works, with the only hitch being that it requires eight fermionic T-dualities and works too well! In short, self-duality is realised using the sixteen standard Killing spinors of pp-waves combined into eight commuting fermionic directions, and as such, it appears to be a symmetry of generic pp-waves. Though we appear to lose information after taking the pp-wave limit making it not possible to infer much about self-duality in the $AdS_4 \times \mathbb{C}\textrm {P}^3$, the obvious silver lining is that we identify a potential new symmetry of pp-waves. We support this claim by studying pp-waves with differing amounts of unbroken supersymmetry and in each case find the same self-duality with identical factors. 

In addition, in this paper we extend the results of \cite{Bakhmatov:2009be} by studying fermionic T-duality with respect to the supernumeracy Killing spinors in the setting of the maximally supersymmetric pp-wave of type IIB. These Killing spinors depend on transverse coordinates and, at least for pp-waves, are somewhat analogous to superconformal supercharges in the Poincar\'e patch. Though we show that one can find consistent equations which may be integrated to give non-trivial examples of fermionic T-duality, we find that fermionic T-duality with respect to these spinors will not recover the original geometry and that there is no self-duality with respect to these spinors. 

\section{Maximally supersymmetric pp-wave in IIB}
In this section, we review some of the analysis presented in \cite{Bakhmatov:2009be} where fermionic T-duality of the maximally supersymmetric pp-wave \cite{IIBpp} in type IIB is discussed. This will allow us to cover the basic recipe for fermionic T-duality and get our bearings by seeing how the requirement of commuting supersymmetries selects Killing spinors and thus fermionic isometries. We quickly review the prescription given in \cite{fermTdual} for performing T-duality. Our gamma matrices conventions which we use throughout this study may be found in the appendix.

Given a type IIB supergravity solution and a Killing spinor that it preserves, the existence of a fermionic isometry requires a pair of sixteen component Weyl spinors of positive chirality $(\e, \hat{\e})$ satisfying \cite{fermTdual}
\be
\label{fermTcond1}
\e \g^{\mu} \e + \hat{\e} \g^{\mu} \hat{\e} = 0,
\ee
where $\mu = 0, 1, \cdots ,9$ and we have dropped spinorial indices. Since, in a Majorana representation, $\g^0$ is simply the identity matrix, it is easy to convince oneself that  $\e$ and $\hat{\e}$ must be complex, otherwise no non-trivial solution exists. Thus, the need for complex Killing spinors in all discussions of fermionic T-duality.

In the case of multiple fermionic isometries, the constraint above generalises to
\be
\label{fermTcond1b}
\e_i \g^{\mu} \e_j + \hat{\e}_i \g^{\mu} \hat{\e}_j = 0,
\ee
where the indices $i,j = 1,...,n$ range over $n$, the number of fermionic T-dualities performed. Once the Killing spinors $(\e_i, \hat{\e}_i)$ are identified, one then computes an auxiliary matrix $\tilde{C}_{ij}$
\bea
\label{parCIIB}
\partial_{\mu} \tilde{C}_{ij} &=& i  \e_i \g^{\mu} \e_j - i\hat{\e}_i \g^{\mu} \hat{\e}_j, \nn
&=& 2 i \e_i \g^{\mu} \e_j .
\eea
In turn, this determines the shift in the dilaton
\be
\label{dilshift} 
\tilde{\phi} = \phi + \frac{1}{2} \sum_{i=1}^{n}(\log \tilde{C})_{ii},
\ee
and an accompanying rearrangement of the fluxes
\be
\frac{i}{16} e^{\tilde{\phi}} \tilde{F} = \frac{i}{16} e^{\phi} F - \tilde{C}^{-1}_{ij} \e_i \otimes \hat{\e}_j.
\ee
Here $F$ is a bispinor  incorporating the RR forms of type IIB supergravity
\be
F^{\a \b} = (\g^{\mu})^{\a \b} F_{\mu} + \frac{1}{3!} (\g^{\mu_1 \mu_2 \mu_3})^{\a \b} F_{\mu_1 \mu_2 \mu_3} + \frac{1}{2} \frac{1}{5!} (\g^{\mu_1 \mu_2 \mu_3 \mu_4 \mu_5})^{\a \b} F_{\mu_1 \mu_2 \mu_3 \mu_4 \mu_5},
\ee
allowing the fluxes to be read off.

Having introduced the nuts and bolts of the transformation, we now review some of the analysis presented in \cite{Bakhmatov:2009be} in the pp-wave setting pertaining to how one selects the pair $(\e, \hat{\e})$. The maximally supersymmetric pp-wave background in type IIB is given by
\begin{subequations}
\be
\label{ppwave-ds}
ds^2 = 2 dx^+ dx^- - \l^2 \d_{\m\n} x^\m x^\n dx^+ dx^+ + \d_{\m\n} dx^\m dx^\n,
\ee
\be
\label{ppwave-RR}
F_{+1234} = 4\l = F_{+5678},
\ee
\end{subequations}
where $\l$ is a constant.

The Killing spinors of this background have been derived in \cite{IIBpp} and in our notation are given by
\be
\label{pp-killing}
\eta = \left(\mathbb{1} - i x^\s \mathbb{A}_\s\right) \left(\cos\frac{\l x^+}{2} \mathbb{1} - i \sin\frac{\l x^+}{2}\mathbb{I}\right) \left(\cos\frac{\l x^+}{2} \mathbb{1} - i \sin\frac{\l x^+}{2}\mathbb{J}\right) \eta_0,
\ee
for an arbitrary $\eta_0$, where $\mathbb{1}$ is a $32 \times 32$ unit matrix, $\mathbb{I} = \G_1\G_2\G_3\G_4$, $\mathbb{J} = \G_5\G_6\G_7\G_8$, and
\be
\label{Amat}
\mathbb{A}_\s = \left\{
\begin{array}{ll}
8\l\, \G_-\, \mathbb{I}\, \G_\s, & \s = 1,2,3,4,\\
8\l\, \G_-\, \mathbb{J}\, \G_\s, & \s = 5,6,7,8.
\end{array}
\right.
\ee

Being Weyl and of positive chirality, we see from our gamma matrices in the appendix that $\eta_0$ has 16 upper components. These 16 components can further be divided into ``A" type and ``B" type Killing spinors in the notation of \cite{Bakhmatov:2009be}. Respectively, these A and B type Killing spinors correspond to Killing spinors in the kernel of $\G^{+}$, often referred to as \textit{standard} Killing spinors as they are common to all pp-waves, and the remaining Killing spinors, dubbed \textit{supernumeracy} Killing spinors.

In terms of components these spinors may be written
\bea
\eta_0^{(A)} = \left( \begin{array}{c} 1 \\ 0 \end{array}\right) \otimes \left( \begin{array}{c} 1 \\ 0 \end{array}\right) \otimes \xi^{(A)}, \nn
\eta_0^{(B)} = \left( \begin{array}{c} 1 \\ 0 \end{array}\right) \otimes \left( \begin{array}{c} 0 \\ 1 \end{array}\right) \otimes \xi^{(B)},
\eea
where $\xi^{(A)}$ and $\xi^{(B)}$ are generic \textit{constant} complex spinors with 8 components. When these are inserted back in (\ref{pp-killing}), one finds that the Killing spinors take the following form
\bea
\label{ABkill}
\eta^{(A)} &=& \left( \begin{array}{c} 1 \\ 0 \end{array}\right) \otimes \left( \begin{array}{c} 1 \\ 0 \end{array}\right) \otimes \left[ \cos \l x^+ \mathbb{1}_8 + i \sin \l x^+ (\s_3 \otimes \s_2 \otimes \s_2)\right] \xi^{(A)}, \nn
\eta^{(B)} &=& \left[\mathbb{1}_{32} - i x^\s \mathbb{A}_{\s} \right] \left( \begin{array}{c} 1 \\ 0 \end{array}\right) \otimes \left( \begin{array}{c} 0 \\ 1 \end{array}\right) \otimes  \xi^{(B)}.
\eea
We see immediately that A type Killing spinors are independent of the transverse coordinates and depend only on $x^+$, whereas the opposite is true for B type Killing spinors. In the next section, we will recognise this as a common feature of pp-waves that arise as a limit of $AdS_4 \times \mathbb{C} \textrm{P}^3$.

Now, in \cite{Bakhmatov:2009be}, the real and imaginary parts of a complex Weyl spinor corresponding to a standard Killing spinor were chosen giving rise to a pair of Majorana-Weyl spinors, which once \textit{complexified}, resulted in a pair satisfying (\ref{fermTcond1}). In each case considered there, $\hat{\e} = i \e$, and thus
\be
\e \g^{\mu} \e + \hat{\e} \g^{\mu} \hat{\e} = \e \g^{\mu} \e - \e \g^{\mu} \e = 0.
\ee
As one is quick to note, the commutation condition is trivially satisfied. To get a better understanding of how one arrives at this choice, we choose to rewrite (\ref{fermTcond1}) in terms of conditions on components coming from the constant Weyl spinor $\eta_0$ appearing in (\ref{pp-killing}).


We begin with the type A (standard) Killing spinors and adopt
\be
\e = \left( \begin{array}{c} 1 \\ 0 \end{array}\right) \otimes \left[ \cos \l x^+ \mathbb{1}_8 + i \sin \l x^+ (\s_3 \otimes \s_2 \otimes \s_2)\right] \xi,
\ee
with a similar hatted expression for $\hat{\e}$. Observe that as the ten-dimensional spinors are of positive chirality, we have dropped the first tensor product resulting in a 16 component spinor. From the gamma matrix decomposition (\ref{gammadec}), we now see that only the $\gamma^0$ and $\gamma^9$ conditions from (\ref{fermTcond1}) are non-trivial and imply the condition
\be
\xi \xi + \hat{\xi} \hat{\xi} = 0.
\ee
We immediately see that a very simple solution to this condition involves $\hat{\xi} = i \xi$, or $\hat{\e} = i \e$. Also, when one is performing numerous fermionic T-dualities using the A type Killing spinors, this condition gets generalised to
\be
\xi_i \xi_j + \hat{\xi}_i \hat{\xi}_j = 0,
\ee
where $i,j = 1,\cdots, n$ with $n$ denoting the number of fermionic T-dualities performed. As we have eight complex components for the spinor $\xi$, we see that we can make eight complex spinors which correspond to the eight commuting fermionic directions noted in \cite{Bakhmatov:2009be} that lead to a self-dual pp-wave. Once the fermionic T-duality condition (\ref{fermTcond1}) is rewritten in terms of the constant spinors, one gets an immediate appreciation for why the choice $\hat{\e} = i \e$, a choice which trivially satisfies the condition, is the natural one to consider.

We next shift focus to the B type (supernumeracy) Killing spinors in (\ref{ABkill}) and adopt the simple choice for the spinor
\be
\label{alpha1}
{\e} = \left[\mathbb{1}_{16} - i x^\s \mathbb{\tilde{A}}_{\s} \right]  \left( \begin{array}{c}  0 \\ 1 \end{array} \right) \otimes \left( \begin{array}{cccccccc} \a_1 & \a_2 & \a_3 & \a_4 & \a_5 &  \a_6 & \a_7 & \a_8 \end{array} \right)^t,
\ee
where $\alpha_i \in \mathbb{C}$ are constants, the superscript $t$ denotes the transpose and we have introduced $\mathbb{\tilde{A}}$ to denote an obvious sixteen-dimensional matrix derived from (\ref{Amat}). Our task now is to find another constant positive chirality  $\hat{\e}$ so that (\ref{fermTcond1}) is satisfied. For a general spinor of the form
\bea
\hat{\e} = \left[\mathbb{1}_{16} - i x^\s \mathbb{\tilde{A}}_{\s} \right]  \left( \begin{array}{c}  0 \\ 1 \end{array} \right) \otimes \left( \begin{array}{cccccccc} \b_1 & \b_2 & \b_3 & \b_4 & \b_5 & \b_6 & \b_7 & \b_8 \end{array} \right)^t,   
\eea
where again $\b_i \in \mathbb{C}$, it can be shown by expanding the condition (\ref{fermTcond1}) in terms of the transverse coordinates, that given $\e$, the only solution for $\hat{\e}$ is with components $\b_i$ satisfying
\be
\b_i = \pm i \a_i, ~~i = 1,..,8.
\ee
Again we see that the simple choice $\hat{\e} = i \e$ emerges when one considers the supernumeracy Killing spinors too. In the next section we explore fermionic T-duality using supernumeracy Killing spinors. 

\subsection{Supernumeracy Killing spinors}
As fermionic T-duality with respect to the supernumeracy Killing spinors has yet to be discussed in the literature, here we ask whether they may be used to generate new backgrounds.

As we have seen above, a simple first choice could involve $\a_1 = 1, \b_1 = i$, with all other components of the spinor zero. However, even with (\ref{fermTcond1}) satisfied, one quickly encounters another problem. Namely, one finds that the equations above are inconsistent in the sense that $\partial_{\pm} \tilde{C} \neq 0$ when $\a_1 = 1, \b_1 = i$, whereas the RHS of (\ref{parCIIB}) is independent of $x^{\pm}$ since the Killing spinors are independent. This obstacle may be overcome by also turning on $\a_2$ and $\b_2$ such that $\a_1 =1, \a_2 = i, \b_1 = i, \b_2 = -1$. This identifies one fermionic direction and the resulting $\tilde{C}$
\be
\tilde{C} = 64 \sqrt{2} i \lambda (x_1 - i x_4)(x_6-ix_7).
\ee
determines the dual geometry by identifying how the dilaton and RR fluxes transform.

Proceeding in this fashion to find extra commuting supersymmetries, one can find four linearly independent solutions to (\ref{fermTcond1b}) with the added requirement that the equation (\ref{parCIIB}) can be integrated. In the absence of the latter condition, one would find eight. The four $\e_i$ may be expressed as
\be
\e_i =   \left[\mathbb{1}_{16} - i x^\s \mathbb{\tilde{A}}_{\s} \right]  \left( \begin{array}{c}  0 \\ 1 \end{array} \right) \otimes \xi_i,
\ee
where
\bea
&& \xi_1=
 \left( \begin{array}{cccc}1 \\   0 \end{array} \right) \otimes \left( \begin{array}{cccc}1 \\   0  \end{array} \right)\otimes \left( \begin{array}{cccc}1 \\   i  \end{array} \right), \;\;\;\;                                      \xi_2=
 \left( \begin{array}{cccc}1 \\   0 \end{array} \right) \otimes \left( \begin{array}{cccc}0 \\   1  \end{array} \right)\otimes \left( \begin{array}{cccc}1 \\   i  \end{array} \right), \cr\cr                                      && \xi_3=
 \left( \begin{array}{cccc} 0 \\   1 \end{array} \right) \otimes \left( \begin{array}{cccc}1 \\   0  \end{array} \right)\otimes \left( \begin{array}{cccc}1 \\   i  \end{array} \right), \;\;\;\;                                      \xi_4= \left( \begin{array}{cccc} 0 \\   1 \end{array} \right) \otimes \left( \begin{array}{cccc}0 \\   1  \end{array} \right)\otimes \left( \begin{array}{cccc}1 \\   i  \end{array} \right). \nonumber
\eea
where in each case $\hat{\e}_i = i \e_i$. Proceeding to calculate $\tilde{C}$, one finds
\be
\tilde{C} = 64\sqrt{2} i \lambda \left( \begin{array}{cccc} z_1z_2  & 0 & -x_2z_2 & x_3z_2 \\  0 & z_1z_2 & x_3z_2 & x_2z_2 \\ -x_2z_2 & x_3z_2& -\bar{z}_1z_2 & 0 \\ x_3z_2 & x_2z_2 & 0 & -\bar{z}_1z_2 \end{array} \right),
\ee
where $z_1=x_1-ix_4$ and $z_2=x_6-ix_7$. Inverting this matrix and contracting to form $\tilde{C}_{ij}^{-1} \e_i \otimes \hat{\e}_j$ leads to a matrix that is not proportional to the spinor bilinear $i e^{\phi} F$. We see that although non-trivial $\tilde{C}$ matrices which presumably lead to involved dual geometries may be constructed, combining commuting fermionic isometries does not lead to any hint of self-duality.

\section{ABJM pp-wave in type IIA}
Shortly after ABJM \cite{ABJM} appeared, the pp-wave limit of the geometry $AdS_4 \times \mathbb{C} \textrm{P}^3$ was determined in \cite{ABJMpp}, in the process making contact with older works on type IIA pp-waves preserving twenty four supersymmetries \cite{benaroiban,N24pp1,N24pp2}. These geometries were originally found by reducing the maximally supersymmetric pp-wave \cite{D11pp} of $D=11$ supergravity. In this section, in the spirit of the work of \cite{IIBpp,D11pp}, we work out the Killing spinors for the ABJM pp-wave limit. A similar treatment with the same conclusion may be found in \cite{N24pp1}.

Adopting the notation of \cite{N24pp2}, the solution may be written \bea
\label{metric}
ds^2 &=& - 2 dx^{+} dx^{-} - A(x^i) (dx^+)^2 + \sum_{i=1}^8 (dx^i)^2, \nn
F_{+123} &=& \mu, \quad F_{+4} = - \frac{\mu}{3},
\eea
where
\be
\label{A}
A(x^i) = \sum_{i=1}^4 \frac{\mu^2}{9} (x^i)^2 + \sum_{i=5}^8 \frac{\mu^2}{36} (x^i)^2.
\ee
It is an easy task to check that this solution satisfies the Einstein equation (\ref{Einstein}), so we only have to concern ourselves with showing that it preserves twenty four supercharges.

We begin by introducing an orthonormal frame
\be
ds^2 = - 2 e^{+} e^{-} + (e^{i})^2,
\ee
where
\bea
e^{+} &=& dx^+, \nn
e^{-} &=& dx^- + \frac{1}{2} A(x^i) dx^+, \nn
e^{i} &=& dx^i.
\eea
Note now that the only non-vanishing component of the spin connection is
\be
\omega^{-i} = \frac{1}{2} \partial_i A(x^i) dx^+,
\ee
 and that with our choice of frame $\delta_{+-} = -1$, so raising and lowering plus and minus indices results in a change of sign.

Using the supersymmetry variations in the appendix, the dilatino variation can be shown to vanish provided
\be
\label{dilatino}
\G^{+} (\G^{1234} \G^{11} - 1) \epsilon = 0.
\ee

From the gravitino variation $\delta \Psi_{M} = (\nabla_M + \Omega_M) \e = 0$, the respective $\Omega$ may be written
\bea
\Omega_- &=& 0, \nn
\Omega_+ &=& - \frac{\mu}{96} I \left[ \G^{-+} (9 - \G^{1234} \G^{11}) + (15-7 \G^{1234} \G^{11})\right], \nn
\Omega_i &=& - \frac{\mu}{6} I \G^{+} \G^{i}, \quad i = 1,2,3,4, \nn
\Omega_i &=& - \frac{\mu}{12}  I \G^{+} \G^{i}, \quad i = 5,6,7,8,
\eea
where $I \equiv \G^{123}$.

After doing some preparatory groundwork, we can now solve the gravitino Killing spinor equation. We start by imposing $\G^{+} \epsilon^{(+)} = 0$, so that the dilatino variation (\ref{dilatino}) is trivially satisfied. Here the superscript on $\e^{(+)}$ simply refers to $\ker \G^+$, the kernel of $\G^{+}$. The Killing spinor equations then reduce to the single equation
\be
\label{diffx}
\frac{d}{d x^{+}}  \e^{(+)} - \frac{\mu}{4} I (1 - \frac{1}{3} \G^{1234} \G^{11}) \e^{(+)} = 0,
\ee
where $\epsilon^{(+)}$ is just a function of $x^+$. Solving for $\e^{(+)}$ one finds
\be
\label{G+ker}
\e^{(+)} = e^{\tfrac{\mu}{6} I x^+} \e^{(+)}_{+} + e^{\tfrac{\mu}{3} I x^+} \e^{(+)}_-,
\ee
where $\e^{(+)}_{\pm}$ are constant spinors satisfying $\G^{1234} \G^{11} \e^{(+)}_{\pm} = \pm \e^{(+)}_{\pm}$.

After working out the form of the standard Killing spinors, we now shift attention to the supernumeracy Killing spinors. Again one easily confirms that they are independent of $x^-$, but need to satisfy (\ref{dilatino}) and
\be
\delta \Psi_{i} = (\partial_i + \Omega_i) \e = 0.
\ee
Now, as $\Omega_i \Omega_j = 0 ~\forall ~i, j$, the supernumeracy Killing spinors may be expressed as
\be
\label{chi}
\e = (1 - x^i \Omega_i ) \chi,
\ee
with $\G^{1234} \G^{11} \chi = \chi$ following from the dilatino variation\footnote{There is a slight subtlety here. $\Omega_i$ for $i=1,\cdots, 4$ anti-commutes with $\G^{1234} \G^{11}$, whereas $\Omega_i$ with $i=5,\cdots,8$ commute. However, as the dilatino variation also has a $\G^+$ term, the $\Omega_i$ terms are killed. }.  Finally, to determine the final form of the supernumeracy Killing spinors, one needs to consider the vanishing of the gravitino variation $\delta \Psi_+ = (\nabla_+ + \Omega_+) \e$. Inserting (\ref{chi}) into $\delta \Psi_+$, after a little manipulation, one finds the equation
\bea
\frac{d }{d x^+} \chi &=& \frac{\mu}{96} I \left( [\G^- \G^+ +1] (9 - \G^{1234} \G^{11}) + (15-7 \G^{1234} \G^{11}) \right) \chi \nn
&-& x^{i} \biggl( \frac{\delta_{ij}}{4 x^{i}} \partial_j A \G^{j+} + \frac{\mu}{96} I [(9 - \G^{1234} \G^{11}) + (15- 7\G^{1234} \G^{11})] \Omega_i \nn &+& \frac{\mu}{96} \Omega_i I [(9 - \G^{1234} \G^{11}) - (15-7 \G^{1234} \G^{11})]\biggr) \chi.
\eea
As in \cite{IIBpp, D11pp}, the terms in the above expression proportional to $x^{i}$ can be shown to vanish using (\ref{A}) and the projector $\G^{1234} \G^{11} \chi = \chi$. One can then further decompose the spinor $\chi = \chi_+ + \chi_-$, where $\G^{\pm} \chi_{\pm} = 0$, leading to two equations
\be
\frac{d}{d x^+} \chi_+ = \frac{\mu}{6} I \chi_+, \quad \frac{d}{d x^+} \chi_- = 0.
\ee
Solving the first equation, one recovers information about the $e^{(+)}_+$ term in (\ref{G+ker}), while the second equation tells us that $\chi_-$ is a constant. This means that the final form of the Killing spinor is
\be
\label{n24killspin}
\e =  e^{\tfrac{\mu}{6} I x^+} \e^{(+)}_{+} + e^{\tfrac{\mu}{3} I x^+} \e^{(+)}_-  +  (1- x^i \Omega_i) \e^{(-)}_+,
\ee
where we have relabelled $\chi_- = \e^{(-)}_+$ to highlight that it is in $\ker \G^-$ and also an eigenspinor of $\G^{1234} \G^{11}$ with eigenvalue $1$. We can quickly confirm that twenty four Killing spinors are preserved: $\e^{(+)} \in \ker \G^+$ correspond to sixteen, with a further eight coming from $\e^{(-)}_+$ as it satisfies a further projection.

\subsection{Fermionic T-duality}
Following observations in section 2, it appears that supernumeracy Killing spinors will not permit self-duality. In addition, in the current setting of the pp-wave limit of ABJM, we also have half the number of supernumeracy Killing spinors, thus making the possibility even more remote. So, here we focus exclusively on the standard Killing spinors.

As we have seen, these are eigenspinors of $\G^{1234} \G^{11}$ and are also in $\ker \G^+$. Defining $\G^{+} = \tfrac{1}{\sqrt{2}} (\G^{9} + \G^0)$, the Killing spinors satisfying $\G^+ \cale = (\G^{1234} \G^{11} -1) \cale = 0$ may be, using the gamma matrices in the appendix, built from a basis of the following spinors 
\bea
\label{spb1} \left.\begin{array}{c} \eta_1 \\ \eta_2 \\ \eta_3 \\ \eta_4 \end{array} \right\}&=& \xi_{+} \otimes \xi_{+} \otimes \Biggl\{ \begin{array}{c} \xi_+ \otimes \biggl\{ \begin{array}{c}\zeta_+ \otimes \zeta_- \\ \zeta_- \otimes \zeta_+ \end{array} \\ \xi_- \otimes \biggl\{ \begin{array}{c}\zeta_+ \otimes \zeta_+ \\ \zeta_- \otimes \zeta_- \end{array} \end{array} \\
\label{spb2} \left.\begin{array}{c} \eta_5 \\ \eta_6 \\ \eta_7 \\ \eta_8 \end{array} \right\}&=& \xi_{-} \otimes \xi_{-} \otimes \Biggl\{ \begin{array}{c} \xi_+ \otimes \biggl\{ \begin{array}{c}\zeta_+ \otimes \zeta_+ \\ \zeta_- \otimes \zeta_- \end{array} \\ \xi_- \otimes \biggl\{ \begin{array}{c}\zeta_+ \otimes \zeta_- \\ \zeta_- \otimes \zeta_+ \end{array}\end{array}
\eea
where $\xi$ and $\zeta$ denote the eigenspinors of $\sigma_3$ and $\sigma_2$ respectively, i.e. $\s_3 \xi_{\pm} = \pm \xi_{\pm}$ and $\s_2 \zeta_{\pm} = \pm \zeta_{\pm}$.  Interchanging the first two spinor products in (\ref{spb1}) and (\ref{spb2}) leads to a basis for Killing spinors satisfying $\G^+ \cale = (\G^{1234} \G^{11} +1) \cale = 0$, which we also record here
\bea
\label{spb3} \left.\begin{array}{c} \eta_1' \\ \eta_2' \\ \eta_3' \\ \eta_4' \end{array} \right\}&=& \xi_{+} \otimes \xi_{+} \otimes \Biggl\{ \begin{array}{c} \xi_+ \otimes \biggl\{ \begin{array}{c}\zeta_+ \otimes \zeta_+ \\ \zeta_- \otimes \zeta_- \end{array} \\ \xi_- \otimes \biggl\{ \begin{array}{c}\zeta_+ \otimes \zeta_- \\ \zeta_- \otimes \zeta_+ \end{array}\end{array}\\
\label{spb4} \left.\begin{array}{c} \eta_5' \\ \eta_6' \\ \eta_7' \\ \eta_8' \end{array} \right\}&=& \xi_{-} \otimes \xi_{-} \otimes \Biggl\{ \begin{array}{c} \xi_+ \otimes \biggl\{ \begin{array}{c}\zeta_+ \otimes \zeta_- \\ \zeta_- \otimes \zeta_+ \end{array} \\ \xi_- \otimes \biggl\{ \begin{array}{c}\zeta_+ \otimes \zeta_+ \\ \zeta_- \otimes \zeta_- \end{array}\end{array} \eea
Note there are a total of 16 basis spinors, 8 for each eigenvalue of $\G^{1234} \G^{11}$.

As explained in \cite{Bakhmatov:2010fp} in the context of type IIA, the condition of the existence of an Abelian fermionic isometry \cite{fermTdual} may be expressed as
\be
\label{fermiso}
\bar{\cale}_i \G^{\mu} \cale_j = 0,
\ee
where $\cale_i$ denotes a 32-component non-chiral complex spinor and we are working in a Majorana-Weyl representation of the gamma matrices with $\bar{\cale} = \cale^{t} C$. The auxiliary function $\tilde{C}$ of fermionic T-duality may then be written
\be
\label{auxC} \partial^{\mu} \tilde{C}_{ij} = i \bar{\cale}_i \G^{\mu} \G^{11} \cale_j.
\ee

Before proceeding to find the Killing spinors satisfying (\ref{fermiso}) which lead to a non-trivial $\tilde{C}$, we digress briefly to explore the options. Adopting the spinor notation $\cale_{\pm} \in \ker \G^{+}$ such that $\G^{1234} \G^{11} \cale_{\pm} = \pm \cale_{\pm}$, one can immediately infer using the projection conditions that
\bea
\label{cond1} &&\bar{\cale}_{\pm} \G^{\mu} \G^{11} \cale_{\pm} = \bar{\cale}_{\pm} \G^{\mu} \cale_{\pm} = 0, \quad \mu = 1,2,3,4, \\
\label{cond2} &&\bar{\cale}_{+} \G^{\mu} \G^{11} \cale_{-} = \bar{\cale}_{+} \G^{\mu} \cale_{-} = 0, \quad \mu = +,-,5,6,7,8.
\eea

Referring to the last section where we have seen that $\cale_{\pm}$ depend only on $x^+$, we can conclude that there is no non-trivial solution to (\ref{auxC}) when one mixes different eigenspinors of $\G^{1234} \G^{11}$. In this case, the only solution is constant $\tilde{C}$. Indeed, one can go further and show that there is no non-trivial fermionic isometry direction with mixed spinors. This can be done by using the explicit basis for the Killing spinors above to show that $\bar{\cale}_{+} \G^{\mu} \G^{11} \cale_{-} = \bar{\cale}_{+} \G^{\mu} \cale_{-} = 0 ~\forall ~\mu$. So, we can ignore this possibility and simply focus on finding fermionic isometry directions that involve either $\cale_+$ or $\cale_-$, not both.

Returning to (\ref{n24killspin}), and confining ourselves to spinors in $\ker \G^{+}$ one recognises that we have two exponentials involving $I$ with different factors. A little experimentation to satisfy the condition (\ref{fermiso}) reveals that the four complex spinors 
\bea
\label{spin1}
\cale^1_0 &=& \eta_1  - \eta_8, \quad
\cale^2_0 = \eta_2  + \eta_7, \nn
\cale^3_0 &=& \eta_3  - \eta_6, \quad
\cale^4_0 = \eta_4  + \eta_5, 
\eea
can be exponentiated as $\cale^i = e^{\tfrac{\mu}{6} I x^+} \cale^i_0$, $i=1,2,3,4$ to give four commuting fermionic directions. Another four complex spinors 
\bea
\cale^5_0 &=& \eta_1'  - \eta_8', \quad
\cale^6_0 = \eta_2'  + \eta_7', \nn
\cale^7_0 &=& \eta_3'  - \eta_6', \quad
\cale^8_0 = \eta_4'  + \eta_5',
\eea
exponentiating as $\cale^i = e^{\tfrac{\mu}{3} I x^+} \cale^i_0$, $i=5,6,7,8$ give us another four commuting complex directions. 

The choice of signs is a result of imposing (\ref{fermiso}). Also, as we have seen above, spinors which have different signs under the projector $\G^{1234} \G^{11}$ do not mix, thus guaranteeing that we have eight commuting fermionic directions. 

Calculating $\tilde{C}$ leads to the following matrix
\be
\label{tildeCN24}
\tilde{C} = \left( \begin{array}{cccccccc} A & -iB & 0 & 0 & 0 & 0 & 0 & 0  \\
-iB & A  & 0 & 0 & 0 & 0 & 0 & 0  \\
0 & 0 & A & -iB & 0 & 0 & 0 & 0  \\
0 & 0 & -iB  & A & 0 & 0 & 0 & 0  \\
0 & 0 & 0 & 0 & -A' & -iB' & 0 & 0  \\
0 & 0 & 0 & 0 & -iB' & -A' & 0 & 0  \\
0 & 0 & 0 & 0 & 0 & 0 & -A' & -iB' \\
0 & 0 & 0 & 0 & 0 & 0 & -iB' & -A'
 \end{array} \right),
\ee
where
\bea
A &=&  \frac{24 \sqrt{2}}{\mu} \cos \left( \frac{\mu x^{+}}{3} \right), \quad B =  \frac{24 \sqrt{2}}{\mu} \sin \left( \frac{\mu x^{+}}{3} \right), \nn
A' &=& \frac{12 \sqrt{2}}{\mu} \cos \left( \frac{2 \mu x^+}{3}\right), \quad B'=\frac{12 \sqrt{2}}{\mu} \sin \left( \frac{2 \mu x^+}{3}\right).
\eea

\subsection{Self-duality}
The transformation of the RR fields under fermionic T-duality \cite{fermTdual} may be written
\be
\label{fluxtransf}
i e^{\tilde{\phi}} \tilde{F}^{\a}_{~ \b} = i e^{\phi} F^{\a}_{~ \b} - 16\tilde{C}_{ij}^{-1} \e^{\a}_i \otimes \hat{\e}_{\b \;j},
\ee
where $i,j$ range over the number of fermionic T-dualities to be performed and $\e_i$ and $\hat{\e}_i$ refer to chiral and anti-chiral components of $\cale_i$ respectively. Above $F^{\a}_{~\b}$ is a bispinor
\be
\label{bispin}
F^{\a}_{~ \b} = \tfrac{1}{2!} (\gamma^{\mu \nu } \bar{c} )^{\a}_{~ \b} F_{\mu \nu} + \tfrac{1}{4!} (\gamma^{\mu \nu \rho \sigma} \bar{c})^{\a}_{~ \b} F_{\mu \nu \rho \sigma},
\ee
where we have taken care with raised and lowered spinor indices through the introduction of $\bar{c}$ - see (\ref{16dblock}).
 Note that the factors in (\ref{fluxtransf}) are the same as those that appeared in \cite{Bakhmatov:2009be}. For self-duality, we simply require that both matrices on the RHS of (\ref{fluxtransf}) be identical up to some constant. For the original solution
\be
i e^{\phi} F^{\a}_{~ \b} = -\sqrt{2} \mu \left( \begin{array}{cc} 0 & 1 \\ 0 & 0  \end{array}\right) \otimes \left[ -\tfrac{1}{3}(\s_1  \otimes \s_2  \otimes 1)  -i (\s_2  \otimes 1_2  \otimes \s_2) \right],
\ee
where the $\bar{c} \equiv - 1_{16}$ term in (\ref{bispin}) is responsible for the overall minus sign.

Now, this 16 by 16 matrix may be recovered  from the eight spinors in (\ref{spin1}) by projecting out the chiral parts ($\e, \hat{\e}$) and contracting with the inverse of $\tilde{C}$. After a calculation, one finds the following relationship
\be
\label{selfdual}
 8 \tilde{C}^{-1}_{ij} \e_i^{\a} \otimes \hat{\e}_{\b\;j} = i e^{\phi} F^{\a}_{~ \b}.
\ee

The corresponding shift in the dilaton (\ref{dilshift}) may be written
\bea
\tilde{\phi} &=& 4 \log \left( \frac{{24}}{\mu} \right).
\eea

The transformation of the RR fields then takes the simple form in terms of this dilaton
\be
\tilde{F}^{\a}_{~\b} = - e^{-\tilde{\phi}} F^{\a}_{~\b}.
\ee

As a result, the original two-form flux and four-form flux are simply rescaled by the constant dilaton $e^{- \tilde{\phi}}$ in precisely the same fashion as was noted for the maximally supersymmetric pp-wave in type IIB \cite{Bakhmatov:2009be}. A quick glance at the only non-trivial Einstein equation (\ref{Einstein}) and one confirms that the dilaton drops out and the Einstein equation is satisfied as before. 

\section{Other pp-waves}
So, now that we have shown that there is something special about the pp-wave that arises as a limit of the geometry $AdS_4 \times \mathbb{C} \textrm{P}^3$, one can reattempt the same calculation with pp-waves not arising as a limit from $AdS_4 \times \mathbb{C} \textrm{P}^3$. We choose to focus on a superposition pp-wave \cite{benaroiban} preserving twenty supercharges and a generic pp-wave preserving sixteen. We begin with the simpler generic pp-wave. 

For simplicity, we retain $F_{+123} = \mu$, but set the two-form flux from section 3 solution to zero\footnote{Through bosonic T-duality, it makes little difference which form we set to zero. }. This then is a solution to the Einstein equation provided the condition
\be
R_{++} = \frac{1}{2} \partial^i \partial_i A(x^i),
\ee
is met for the metric (\ref{metric}).

We are not interested in the specific form of $A(x^i)$, but it is clear from the form of the dilatino variation that no supernumeracy Killing spinors are preserved. The Killing spinor equation for the standard Killing spinors is then solved for $\cale \in \ker \G^+$, leading to
\be
\label{n16sol}
\cale = e^{\frac{\mu}{4} I x^+} \cale_0,
\ee
where again $\cale_0$ is a constant spinor and $I \equiv \G^{123}$. Using the same basis of  spinors as the previous section (\ref{spin1}), except this time not going to the bother of decomposing in terms of eigenspinors of $\G^{1234} \G^{11}$, one can determine $\tilde{C}$ 
\be
\label{tildeCN16}
\tilde{C} = \left( \begin{array}{cccccccc} A & -iB & 0 & 0 & 0 & 0 & 0 & 0  \\
-iB & A  & 0 & 0 & 0 & 0 & 0 & 0  \\
0 & 0 & A & -iB & 0 & 0 & 0 & 0  \\
0 & 0 & -iB  & A & 0 & 0 & 0 & 0  \\
0 & 0 & 0 & 0 & -A & -iB & 0 & 0  \\
0 & 0 & 0 & 0 & -iB & -A & 0 & 0  \\
0 & 0 & 0 & 0 & 0 & 0 & -A & -iB \\
0 & 0 & 0 & 0 & 0 & 0 & -iB & -A
 \end{array} \right),
\ee
where now 
\be
A = \frac{16 \sqrt{2}}{\mu} \cos \left( \frac{\mu x^{+}}{2} \right), \quad B = \frac{16 \sqrt{2}}{\mu} \sin \left( \frac{\mu x^{+}}{2} \right). 
\ee
Inverting $\tilde{C}$ and contracting with chiral spinors, one finds the same result as (\ref{selfdual}) with an accompanying constant shift in the dilaton.

A less trivial example is provided by a pp-wave with twenty unbroken supersymmetries. As explained in  \cite{benaroiban}, one can superpose two twenty four supercharge pp-waves to get this solution. A flux choice realising this configuration may be written 
\bea
F_{+123} &=& \mu, \quad F_{+4} = - \frac{\mu}{3}, \nn
F_{+456} &=& \lambda, \quad F_{+1} =  \frac{\lambda}{3}.
\eea
Referring to the metric ansatz (\ref{metric}), the new $\Omega$ for this solution are
\bea
\Omega_- &=& 0, \nn
\Omega_+ &=& - \frac{\mu}{96} I \left[ \G^{-+} (9 - \G^{1234} \G^{11}) + (15-7 \G^{1234} \G^{11})\right] \nn &-& \frac{\lambda}{96} J \left[ \G^{-+} (9 - \G^{1456} \G^{11}) + (15-7 \G^{1456} \G^{11})\right], \nn
\Omega_i &=& - \frac{( \mu I + \lambda J)}{6}\G^{+} \G^{i}, \quad i = 1,4 \nn
\Omega_i &=& - \frac{(2 \mu I + \lambda J)}{12}\G^{+} \G^{i}, \quad i = 2,3 \nn
\Omega_i &=& - \frac{(\mu I + 2\lambda J)}{12} \G^{+} \G^{i}, \quad i = 5,6 \nn
\Omega_i &=& - \frac{(\mu I + \lambda J)}{12} \G^{+} \G^{i}, \quad i = 7,8.
\eea
From the supersymmetry variations, the $A(x^i)$ required to complete the solution may be worked out and may be expressed as follows 
\bea
A(x^i) &=& - \frac{1}{9} (\mu^2 + \lambda^2)(x_1^2 + x_4^2) - \frac{1}{36} (4 \mu^2 + \lambda^2)(x_2^2 + x_3^2) \nn &-& \frac{1}{36} (\mu^2 + 4 \lambda^2)(x_5^2 + x_6^2) - \frac{1}{36} (\mu^2 + \lambda^2)(x_7^2 + x_8^2). 
\eea
As a quick check of consistency, we see that setting $\lambda = 0$ we recover the ABJM pp-wave. It is also not difficult to show that the Einstein equation (\ref{Einstein}) is satisfied. 

A glance at the dilatino variation now confirms that in addition to the original projector $\G^{1234} \G^{11} \e = \e$, we also have the additional commuting projector $\G^{1456} \G^{11} \e = \e$. As a result, the supernumeracy Killing spinors are further cut from eight to four. Solving for the standard Killing spinors we meet the differential equation
\be
\partial_{+} \e = \frac{\mu}{4} I (1 - \frac{1}{3} \G^{1234} \G^{11}) \e + \frac{\lambda}{4} J (1 - \frac{1}{3} \G^{1456} \G^{11})\e,
\ee
where $I$ is the same as before in (\ref{diffx}), but now $J \equiv \G^{456}$. The solution for the standard Killing spinors will then read
\bea
\label{n20spin}
\cale &=& e^{ \frac{\mu I + \lambda J}{6} x^+} \cale^{++}_0 +  e^{ \frac{\mu I + \lambda J}{3} x^+} \cale^{--}_0 +   e^{ \frac{\mu I + 2\lambda J}{6} x^+} \cale^{+-}_0 + e^{ \frac{2 \mu I + \lambda J}{6} x^+} \cale^{-+}_0,
\eea
where superscripts now refer to eigenspinors of the two projectors $\G^{1234} \G^{11}$ and $\G^{1456} \G^{11}$.

Proceeding, one can use the following spinors 
\bea
\cale_1 &= & \eta_1 - i \eta_3 +\eta_6 - i \eta_8 , \nn
\cale_2 &=&  \eta_2 + i \eta_4 +\eta_5+i \eta_7, \nn
\cale_3 &=& \eta_1'  + i \eta_3' + \eta_2' - i \eta_4', \nn
\cale_4 &=& \eta_5'-i \eta_7' +  \eta_6' + i \eta_8' , \nn
\cale_5 &=& \eta_1 + i \eta_3 + \eta_2 - i \eta_4, \nn 
\cale_6 &=& \eta_5-i \eta_7 + \eta_6 + i \eta_8, \nn
\cale_7 &=& \eta_1' - i \eta_3' + \eta_2' + i \eta_4', \nn 
\cale_8 &=&\eta_5'+i \eta_7' + \eta_6' - i \eta_8',
\eea
where $\eta_i, \eta_i'$ refer to the spinors we introduced earlier, to find a manageable expression for $\tilde{C}$, which when inverted and contracted leads again to our now familiar self-duality relationship. 
 
So, despite starting off with a result that may appear unique for the pp-wave arising as a limit of $AdS_4 \times \mathbb{C} \textrm{P}^3$, it is clear that there are other examples in the class of pp-waves self-dual under fermionic T-duality. It is also not difficult to see how this result manifests itself.

In all cases, the trigonometric functions of $x^{+}$ that appear in all the expressions for the standard Killing spinors drop out after inverting and contracting $\tilde{C}$. The fact that one is using eight complex Killing spinors also allows one to form a basis for the sixteen component Killing spinors in the kernel of $\G^{+}$. As we have seen, with a varying number of supernumeracy Killing spinors, we find additional projection conditions leading to a subdivision of this basis into smaller bases of Killing spinors honouring the projection conditions. As we are not aware of any exhaustive classification of IIA pp-waves, though steps in that direction in the simpler setting of $D=11$ supergravity may be found in \cite{Cvetic:2002hi,Cvetic:2002si,Gauntlett:2002cs}, we are unable to show this symmetry for all pp-waves. Despite this limitation, evidence so far for the existence of this symmetry in a host of examples  suggests that it is quite general.

\section{Discussion}
In this paper we have studied fermionic T-duality in type IIA pp-waves including the pp-wave limit of $AdS_4 \times \mathbb{C} \textrm{P}^3$. We show that pp-waves preserving different amounts of supersymmetry are self-dual with respect to eight commuting fermionic T-dualities performed with respect to standard Killing spinors, thus generalising the observation in \cite{Bakhmatov:2009be} that the type IIB maximally supersymmetric pp-wave is self-dual. As these Killing spinors are common to all pp-waves, we conjecture that self-duality under fermionic T-duality is a symmetry of all pp-waves. 

We have also explored fermionic T-duality with respect to the supernumeracy Killing spinors for the maximally supersymmetric pp-wave in type IIB. We show that one can find non-trivial auxiliary matrices, but that fermionic T-duality with respect to commuting fermionic directions does not lead to a self-dual configuration. One may expect an analogous statement for fermionic T-duality with respect to superconformal Killing spinors in $AdS_5 \times S^5$. 

Though we have not gleaned any hints into how self-duality may work in the setting of $AdS_4 \times \mathbb{C} \textrm{P}^3$, as there from superconformal algebra arguments one would expect six commuting fermionic T-dualities, while in the pp-wave we need eight, this remains an interesting challenge. It is certainly surprising that in addition to a singularity in the dilaton shift coming from fermionic T-duality, a similar singularity as a result of three commuting bosonic T-dualities on $\mathbb{C} \textrm{P}^3$ arises \cite{Bakhmatov:2010fp}. It would be nice to get a deeper understanding of this effect and any other potential obstacles to self-duality in $AdS_4 \times \mathbb{C} \textrm{P}^3$ as the evidence of the existence of these symmetries from scattering amplitude studies is quite strong. 

Finally, in tandem with recent developments in our understanding of non-Abelian T-duality in coset geometries \cite{Sfetsos:2010uq,Lozano:2011kb}, it may also be interesting to study fermionic T-duality with respect to non-Abelian fermionic T-dualities.

\section{Acknowledgements}
We would like to extend our gratitude to the following for thought-provoking discussion: David Berman, James Drummond, Stefan Hohenegger, Seung-joon Hyun, Sangmin Lee, Juan Maldacena, Tristan McLoughlin and George Papadopoulos. In particular we would like to thank the Seoul fermionic T-duality study group, namely Chang-Young Ee, Hiroaki Nakajima and Hyeonjoon Shin, for constructive feedback. E\'OC is grateful to the Simons Summer workshop 2011 and the Simons Center for Geometry and Physics for hospitality while some of this work took shape. The research of IB is supported by Westfield Trust scholarship.
\appendix

\section{Conventions}
Throughout this work we employ the real representation for the ten-dimensional gamma matrices appearing in the Clifford algebra $Cl(9,1)$. We choose our gamma matrices to be
\be
\label{gammadec}
\G^0 = i\sigma_2 \otimes \mathbb{1}_{16}, \quad \G^i = \sigma_1 \otimes \S^i,
\ee
where we further decompose
\be
\begin{array}{ccccccc}
\S^1 = \sigma_2 & \otimes & \sigma_2 & \otimes & \sigma_2 & \otimes & \sigma_2, \\
\S^2 = \sigma_2 & \otimes & 1             & \otimes & \sigma_1 & \otimes & \sigma_2, \\
\S^3 = \sigma_2 & \otimes & 1             & \otimes & \sigma_3 & \otimes & \sigma_2, \\
\S^4 = \sigma_2 & \otimes & \sigma_1 & \otimes & \sigma_2 & \otimes & 1,\\
\S^5 = \sigma_2 & \otimes & \sigma_3 & \otimes & \sigma_2 & \otimes & 1,\\
\S^6 = \sigma_2 & \otimes & \sigma_2 & \otimes & 1             & \otimes & \sigma_1, \\
\S^7 = \sigma_2 & \otimes & \sigma_2 & \otimes & 1             & \otimes & \sigma_3, \\
\S^8 = \sigma_1 & \otimes & 1             & \otimes & 1              & \otimes & 1, \\
\S^9 = \sigma_3 & \otimes &1  &\otimes & 1 & \otimes & 1.
\end{array}
\ee
Observe here that $\S^9 = \S^1 \cdots \S^8$. Our gamma matrices may be written in terms of 16 dimensional blocks as
\be
\label{16dblock}
\G^{\mu} = \left( \begin{array}{cc} 0 & (\g^{\mu})^{\a \b} \\ \g^{\mu}_{\a \b} & 0 \end{array}\right), ~~C = \left( \begin{array}{cc} 0 & c_{\a}^{~\b} \\  \bar{c}^{\a}_{~\b} & 0 \end{array}\right),~~\G^{11} = \left( \begin{array}{cc} \delta^{\a}_{~ \b} & 0 \\ 0 & \delta_{\a}^{~\b} \end{array}\right),
\ee
where the indices take values $\a, \b = 1,\cdots 16$.

Under the ten-dimensional chirality operator $ \G^{11} = \G^{0} \cdots \G^9  = \s_3 \otimes \mathbb{1}_{16}$, we have two inequivalent 16 component Weyl spinors $\psi_{\pm}$ satisfying $\G^{11} \psi_{\pm} = \pm \psi_{\pm}$. Working in a Majorana representation where $C = \G^0$, and
\bea
C \G^{\m} C^{-1} = - \G^{\m t}, \nn
C^t = - C,
\eea
we see that further imposing the Majorana condition on $\psi_{\pm}$ results in them being real.

In string frame the type IIA supersymmetry conditions may be written as
\bea
\delta \lambda&=&\biggl\{
-\frac{1}{2}\G^M\nabla_M\phi
+\frac{3 e^{\phi}}{16}F^{(2)}_{MN}\G^{MN}\G_{11}
+\frac{1}{24}H^{(3)}_{MNP}\G^{MNP}\G_{11} \nn
&-&\frac{e^{\phi}}{192}F^{(4)}_{MNPQ}\G^{MNPQ}
\biggr\}\e~, \\
\delta \Psi_M &=& \biggl\{ \nabla_M - \frac{1}{8} \Gamma_{M}^{~N} \partial_{N} \phi
-\frac{e^{\phi}}{64}
F^{(2)}_{NP}(\G_M{}^{NP}-14\delta_M{}^N\G^P)\G_{11}\nn\\
&+&\frac{1}{96}H^{(3)}_{NPQ}(\G_M{}^{NPQ}-9\delta_M{}^N\G^{PQ})\G_{11}
+\frac{e^{\phi}}{256}F^{(4)}_{NPQR}(\G_M{}^{NPQR}
-\frac{20}{3}\delta_M{}^N\G^{PQR}) \biggr\} \e. \nonumber
\eea

In addition, any pp-wave solution of type IIA has to satisfy the following equation
\be
\label{Einstein}
R_{++} + 2 \nabla_{+} \nabla_+ \Phi = e^{2 \Phi} \left[ \frac{1}{2} F_{+ \s} F^{~\s}_ {+} + \frac{1}{12} F_{+ \s_1 \s_2 \s_3} F^{~\s_1 \s_2 \s_3}_{+} \right].
\ee


\begin{thebibliography}{99}
\bibitem{Alday}
  L.~F.~Alday, J.~Maldacena,
  ``Comments on gluon scattering amplitudes via AdS/CFT,''
  JHEP {\bf 0711}, 068 (2007).
  [arXiv:0710.1060 [hep-th]].
\bibitem{Drummond:2007aua}
  G.~P.~Korchemsky, J.~M.~Drummond, E.~Sokatchev,
  ``Conformal properties of four-gluon planar amplitudes and Wilson loops,''
  Nucl.\ Phys.\  {\bf B795 } (2008)  385-408.
  [arXiv:0707.0243 [hep-th]].
  \bibitem{Brandhuber:2007yx}
  A.~Brandhuber, P.~Heslop, G.~Travaglini,
  ``MHV amplitudes in N=4 super Yang-Mills and Wilson loops,''
  Nucl.\ Phys.\  {\bf B794}, 231-243 (2008).
  [arXiv:0707.1153 [hep-th]].
\bibitem{fermTdual}
  N.~Berkovits, J.~Maldacena,
  ``Fermionic T-Duality, Dual Superconformal Symmetry, and the Amplitude/Wilson Loop Connection,''
  JHEP {\bf 0809}, 062 (2008).
  [arXiv:0807.3196 [hep-th]].
\bibitem{Beisert}
  N.~Beisert, R.~Ricci, A.~A.~Tseytlin, M.~Wolf,
  ``Dual Superconformal Symmetry from AdS(5) x S**5 Superstring Integrability,''
  Phys.\ Rev.\  {\bf D78}, 126004 (2008).
  [arXiv:0807.3228 [hep-th]].
\bibitem{Buscher}
  T.~H.~Buscher,
  ``A Symmetry of the String Background Field Equations,''
  Phys.\ Lett.\  {\bf B194}, 59 (1987);
  T.~H.~Buscher,
  ``Path Integral Derivation of Quantum Duality in Nonlinear Sigma Models,''
  Phys.\ Lett.\  {\bf B201}, 466 (1988).
\bibitem{de la Ossa:1992vc}
  X.~C.~de la Ossa and F.~Quevedo,
  ``Duality symmetries from nonAbelian isometries in string theory,''
  Nucl.\ Phys.\  B {\bf 403} (1993) 377
  [arXiv:hep-th/9210021].
\bibitem{Sfetsos:2010uq}
  K.~Sfetsos, D.~C.~Thompson,
  ``On non-abelian T-dual geometries with Ramond fluxes,''
  Nucl.\ Phys.\  {\bf B846}, 21-42 (2011).
  [arXiv:1012.1320 [hep-th]].
\bibitem{Lozano:2011kb}
  Y.~Lozano, E.~O.~Colgain, K.~Sfetsos, D.~C.~Thompson,
  ``Non-abelian T-duality, Ramond Fields and Coset Geometries,''
  JHEP {\bf 1106}, 106 (2011).
  [arXiv:1104.5196 [hep-th]].
\bibitem{Adam:2009kt}
  I.~Adam, A.~Dekel, Y.~Oz,
  ``On Integrable Backgrounds Self-dual under Fermionic T-duality,''
  JHEP {\bf 0904 } (2009)  120.
  [arXiv:0902.3805 [hep-th]].
\bibitem{Fre:2009ki}
  P.~Fre, P.~A.~Grassi, L.~Sommovigo, M.~Trigiante,
  ``Theory of Superdualities and the Orthosymplectic Supergroup,''
  Nucl.\ Phys.\  {\bf B825}, 177-202 (2010).
  [arXiv:0906.2510 [hep-th]].
\bibitem{Hao:2009hw}
  C.~-g.~Hao, B.~Chen, X.~-c.~Song,
  ``On Fermionic T-duality of Sigma modes on AdS backgrounds,''
  JHEP {\bf 0912 } (2009)  051.
  [arXiv:0909.5485 [hep-th]].
\bibitem{Dekel:2011qw}
  A.~Dekel, Y.~Oz,
  ``Self-Duality of Green-Schwarz Sigma-Models,''
  JHEP {\bf 1103}, 117 (2011).
  [arXiv:1101.0400 [hep-th]].
\bibitem{ChangYoung:2011rs}
  E.~Chang-Young, H.~Nakajima, H.~Shin,
  ``Fermionic T-duality and Morita Equivalence,''
  JHEP {\bf 1106}, 002 (2011).
  [arXiv:1101.0473 [hep-th]].
\bibitem{Sfetsos:2010xa}
  K.~Sfetsos, K.~Siampos, D.~C.~Thompson,
  ``Canonical pure spinor (Fermionic) T-duality,''
  Class.\ Quant.\ Grav.\  {\bf 28}, 055010 (2011).
  [arXiv:1007.5142 [hep-th]].
\bibitem{Nikolic:2011ps}
  B.~Nikolic, B.~Sazdovic,
  ``Fermionic T-duality and momenta noncommutativity,''
  [arXiv:1103.4520 [hep-th]].
\bibitem{Grassi:2011zf}
  P.~A.~Grassi, A.~Mezzalira,
  ``Aspects of Quantum Fermionic T-duality,''
  JHEP {\bf 1105 } (2011)  019.
  [arXiv:1101.5969 [hep-th]].  
\bibitem{Bargheer:2010hn}
  T.~Bargheer, F.~Loebbert, C.~Meneghelli,
  ``Symmetries of Tree-level Scattering Amplitudes in N=6 Superconformal Chern-Simons Theory,''
  Phys.\ Rev.\  {\bf D82}, 045016 (2010).
  [arXiv:1003.6120 [hep-th]].
\bibitem{Lee:2010du}
  S.~Lee,
  ``Yangian Invariant Scattering Amplitudes in Supersymmetric Chern-Simons Theory,''
  Phys.\ Rev.\ Lett.\  {\bf 105}, 151603 (2010).
  [arXiv:1007.4772 [hep-th]].  
\bibitem{Huang:2010qy}
  Y.~-t.~Huang, A.~E.~Lipstein,
  ``Dual Superconformal Symmetry of N=6 Chern-Simons Theory,''
  JHEP {\bf 1011}, 076 (2010).
  [arXiv:1008.0041 [hep-th]].
 \bibitem{Gang:2010gy}
  D.~Gang, Y.~-t.~Huang, E.~Koh, S.~Lee, A.~E.~Lipstein,
  ``Tree-level Recursion Relation and Dual Superconformal Symmetry of the ABJM Theory,''
  JHEP {\bf 1103}, 116 (2011).
  [arXiv:1012.5032 [hep-th]].
\bibitem{BCFW}
  R.~Britto, F.~Cachazo, B.~Feng and E.~Witten,
  ``Direct Proof Of Tree-Level Recursion Relation In Yang-Mills Theory,''
  Phys.\ Rev.\ Lett.\  {\bf 94} (2005) 181602
  [arXiv:hep-th/0501052].
\bibitem{Bianchi:2011rn}
  M.~S.~Bianchi, M.~Leoni, A.~Mauri, S.~Penati, C.~A.~Ratti, A.~Santambrogio,
  ``From Correlators to Wilson Loops in Chern-Simons Matter Theories,''
  JHEP {\bf 1106}, 118 (2011).
  [arXiv:1103.3675 [hep-th]].
 \bibitem{Bianchi:2011dg}
  M.~S.~Bianchi, M.~Leoni, A.~Mauri, S.~Penati, A.~Santambrogio,
  ``Scattering Amplitudes/Wilson Loop Duality In ABJM Theory,''
   [arXiv:1107.3139 [hep-th]].
\bibitem{Chen:2011vv}
  W.~-M.~Chen, Y.~-t.~Huang,
  ``Dualities for Loop Amplitudes of N=6 Chern-Simons Matter Theory,''
  [arXiv:1107.2710 [hep-th]].
\bibitem{ABJM}
  O.~Aharony, O.~Bergman, D.~L.~Jafferis, J.~Maldacena,
  ``N=6 superconformal Chern-Simons-matter theories, M2-branes and their gravity duals,''
  JHEP {\bf 0810}, 091 (2008).
  [arXiv:0806.1218 [hep-th]].
 \bibitem{Grassi:2009yj}
  P.~A.~Grassi, D.~Sorokin, L.~Wulff,
  ``Simplifying superstring and D-brane actions in $AdS_4 x CP^3$ superbackground,''
  JHEP {\bf 0908}, 060 (2009).
  [arXiv:0903.5407 [hep-th]].
\bibitem{Adam:2010hh}
  I.~Adam, A.~Dekel, Y.~Oz,
  ``On the fermionic T-duality of the $AdS_4 x CP^3$ sigma-model,''
  JHEP {\bf 1010}, 110 (2010).
  [arXiv:1008.0649 [hep-th]].
\bibitem{Bakhmatov:2010fp}
  I.~Bakhmatov,
  ``On AdS$_4$ x CP$^3$ T-duality,''
  Nucl.\ Phys.\  {\bf B847}, 38-53 (2011).
  [arXiv:1011.0985 [hep-th]].
\bibitem{Bakhmatov:2009be}
  I.~Bakhmatov, D.~S.~Berman,
  ``Exploring Fermionic T-duality,''
  Nucl.\ Phys.\  {\bf B832}, 89-108 (2010).
  [arXiv:0912.3657 [hep-th]].
\bibitem{ABJMpp}
  T.~Nishioka, T.~Takayanagi,
  ``On Type IIA Penrose Limit and N=6 Chern-Simons Theories,''
  JHEP {\bf 0808}, 001 (2008).
  [arXiv:0806.3391 [hep-th]].
\bibitem{IIBpp}
  M.~Blau, J.~M.~Figueroa-O'Farrill, C.~Hull, G.~Papadopoulos,
  ``A New maximally supersymmetric background of IIB superstring theory,''
  JHEP {\bf 0201}, 047 (2002).
  [hep-th/0110242].
\bibitem{benaroiban}
  I.~Bena, R.~Roiban,
  ``Supergravity pp wave solutions with twenty eight supercharges and twenty four supercharges,''
  Phys.\ Rev.\  {\bf D67}, 125014 (2003).
  [hep-th/0206195].
\bibitem{N24pp1}
  K.~Sugiyama, K.~Yoshida,
  ``Type IIA string and matrix string on PP wave,''
  Nucl.\ Phys.\  {\bf B644}, 128-150 (2002).
  [hep-th/0208029].
\bibitem{N24pp2}
  S.~-j.~Hyun, H.~-j.~Shin,
  ``N=(4,4) type 2A string theory on PP wave background,''
  JHEP {\bf 0210}, 070 (2002).
  [hep-th/0208074].
\bibitem{D11pp}
  J.~M.~Figueroa-O'Farrill, G.~Papadopoulos,
  ``Homogeneous fluxes, branes and a maximally supersymmetric solution of M theory,''
  JHEP {\bf 0108}, 036 (2001).
  [hep-th/0105308].
\bibitem{Cvetic:2002hi}
  M.~Cvetic, H.~Lu, C.~N.~Pope,
  ``Penrose limits, PP waves and deformed M2 branes,''
  Phys.\ Rev.\  {\bf D69}, 046003 (2004).
  [hep-th/0203082].
\bibitem{Cvetic:2002si}
  M.~Cvetic, H.~Lu, C.~N.~Pope,
  ``M theory p p waves, Penrose limits and supernumerary supersymmetries,''
  Nucl.\ Phys.\  {\bf B644}, 65-84 (2002).
  [hep-th/0203229].
\bibitem{Gauntlett:2002cs}
  J.~P.~Gauntlett, C.~M.~Hull,
  ``Pp-waves in 11 dimensions with extra supersymmetry,''
  JHEP {\bf 0206}, 013 (2002).
  [hep-th/0203255].
\end{thebibliography}
\end{document}